\documentstyle[12pt,aaspp4]{article}

\tighten

\def\xbf{{\bf x}}

\def\thbf{{\bf \theta}}

\def\mubar{\overline{\mu}}

\def\wtheta{{\omega(\theta)}}
\def\b450{B_{450}}
\def\v606{V_{606}}
\def\i814{I_{814}}

\def\mathrm{\rm}
\def\simless{\lesssim}
\def\simgreat{\gtrsim}

\begin{document}

\slugcomment{Submitted to ApJ}

\lefthead{Vogeley}
\righthead{Extragalactic Background Fluctuations}

\title{Fluctuations in the Extragalactic Background Light:\\
Analysis of the Hubble Deep Field\altaffilmark{1}}

\author{Michael S. Vogeley\altaffilmark{2}}

\affil{
Princeton University Observatory\\
Princeton, NJ 08544-1001\\
vogeley@astro.princeton.edu
}

\altaffiltext{1}{
Based on observations with the NASA/ESA {\it Hubble Space Telescope}, obtained
at the Space Telescope Science Institute, which is operated by AURA, under
NASA contract NAS 5-26555
}

\altaffiltext{2}{
Hubble Fellow}

\begin{abstract}

Statistical analysis of the unresolved light in the Hubble Deep Field
(HDF) strongly constrains possible sources of the optical
Extragalactic Background Light (EBL).  This constraint is crucial for
determining the spectrum of the EBL because reported upper limits on
the optical EBL are several times larger than the surface brightness
from detected galaxies, suggesting the possibility of
additional galaxy populations.  To test for the statistical signature
of previously undetected sources, we estimate the auto, cross, and
color correlations of the ``sky'' in the HDF that remains after
masking objects brighter than $\i814=30$mag.  Auto and cross
correlations of surface brightness in the $\v606$ and $\i814$
bandpasses are well-fitted by $\wtheta \sim
10^{-6}(\theta/1'')^{-0.6}$ up to $10''$.  Probable contributions of
several instrumental systematics ensure that these correlations are
firm upper limits on the true EBL fluctuations.  This measurement
yields the most stringent limits to date on small-scale structure in
the night sky; analysis of shallower imaging would be dominated by
galaxies now detected by the HDF.

Unless there is a truly uniform optical background, the mean EBL is
likely to be within a small fraction of the surface brightness from
detected galaxies.  No currently plausible sources of additional EBL
satisfy the constraints that they (1) would not have already been
detected, (2) contribute EBL comparable to that from detected
galaxies, and (3) do not produce EBL fluctuations in excess of the
upper limits set by correlations in the HDF.  These constraints admit
only a confusion-limited population of extremely low surface
brightness objects that is disjoint from the parameter space of all
detected galaxies.  Extrapolation of detected galaxy counts to zero
flux would add only a few percent to the EBL.  Diffuse intergalactic
light clustered similarly to faint galaxies could explain some of the
observed correlations but would contribute at most a few $\times 10\%$
to the mean EBL.

\end{abstract}


\section{INTRODUCTION}

The Extragalactic Background Light (EBL) is the surface brightness of
the night sky from all extragalactic sources, integrated out to the
first epoch of star formation.  The ultimate goal of studying the EBL
is to complete the resolution of Olbers's paradox by measuring the
spectrum of the EBL, identifying the sources that contribute to it,
and accounting for their evolution.  Constraints on the optical EBL
are particularly important because these bandpasses are sensitive to
rest-frame UV flux from galaxies at $z>1$, when we now believe the
bulk of star formation in the universe occurred (Madau et al. 1996;
Connolly et al. 1997).  Thus, measurement of the spectrum of the
optical EBL is a critical test for models of the star formation
history of the universe (see, e.g., Whitrow \& Yallop 1965; Partridge
\& Peebles 1967a,b; Fall, Charlot, \& Pei 1996; Vaisanen 1996; Madau
et al. 1997).  Models for the sources of the EBL must also correctly
predict statistical fluctuations in the EBL, which reflect the surface
brightness profiles and clustering of these sources.  [For an
extensive review on the optical EBL that predates the HDF, see Tyson
1995.]

The average surface brightness of the sky from detected galaxies sets a
firm lower limit on the mean EBL, $\mubar_{EBL} \ge \mubar_{galaxies}$
(here and throughout this paper, $\mubar$ is an average surface
brightness).  Tremendous progress has been made in detecting faint
galaxies, through deep surveys such as the Hubble Deep Field (Williams
et al. 1996) and the near-IR survey by Djorgovski et al. (1995), which
reach as faint as $V=29.5$mag and $K=24$mag, respectively.  However,
it is unclear what fraction of the mean optical EBL the detected
galaxies comprise, because the uncertainties in the mean level of the
optical EBL are larger than the surface brightness contributed by
these galaxies.  In addition, the systematic bias of object detection
against relatively diffuse, low surface brightness objects leaves open
the possibility that we may have missed a substantial fraction of the
optical luminosity density in the universe (Disney 1976; McGaugh et al. 1995;
Dalcanton 1997).

Figure 1 shows various observational constraints on the mean EBL.
The inset box plots only the optical limits, using linear axes.
Solid symbols indicate lower limits on the EBL in several bandpasses
from the average sky brightness contributed by detected galaxies,
$\mubar_{galaxies}$.  In optical bandpasses, these lower limits
include galaxy counts from the HDF and brighter counts from previous
surveys (compilation by Pozzetti et al. 1997).  UV galaxy counts are
from Milliard et al. (1992).  Open circles with $1\sigma$
uncertainties show a recent detection of the mean optical EBL
(Bernstein 1997; Bernstein, Freedman, \& Madore 1998).  Arrows are
upper limits from various experiments (Paresce 1990; Toller 1983;
Dube, Wickes, \& Wilkinson 1977, 1979; Hauser 1996), where the upper
bound in each case is set primarily by uncertainties in foreground
subtraction.  Other bounds on the optical EBL (not plotted here)
include experiments by Roach \& Smith (1968), Lillie (1968), Spinrad
\& Stone (1978), Boughn \& Kuhn (1986), and Mattila (1990).  Precise
measurement of the mean level of the EBL is extremely difficult
because several foreground sources, particularly Zodiacal light, are
at least an order of magnitude brighter than the mean optical EBL.

At optical wavelengths, uncertainties in the mean EBL leave room for
considerable surface brightness above the lower limits set by galaxy
detections.  The $2\sigma$ uncertainties of the optical EBL detection
by Bernstein et al.  span the range from $\mubar_{EBL}\sim 0$ to
$\mubar_{EBL}\sim 5\mubar_{galaxies}$.
The gap between previous upper
limits on the EBL and the galaxy surface brightness is of order
$4-10$.  These uncertainties prompt us to examine the plausibility of
an additional contribution to the EBL that is comparable to the
surface brightness from detected galaxies with magnitude $V<29.5$mag.
Could this flux arise from fainter galaxies?  The surface brightness
per magnitude depends on the number counts as $d I_{\nu}/dm =
(dN/dm)(I_{gal}(m))\propto 10^{(\alpha-0.4) m}$.  If we
extrapolate the galaxy number counts to infinitely faint limits,
assuming the logarithmic count slope $\alpha=0.2$ at the HDF limit,
then all galaxies with $V>29.5$mag would contribute only an additional
$1.6\%$ to the EBL. If, strangely, the number counts turn up to
$\alpha=0.3$ beyond $V=29.5$mag, then this contribution rises to
$3.3\%$. Thus, it seems unlikely that galaxies fainter than the HDF
limit contribute significant flux. However, this does not rule out
objects that evade direct detection, e.g., galaxies with total
magnitude $V < 29.5$ and very low central surface brightness, or a
diffuse intergalactic component.

If a substantial fraction of the EBL resides in sources that now lie
below the detection limits of deep optical imaging, this population
might be inferred from its statistical signature in the object-masked
``sky.''  The internal profiles of sources and clustering among them
will cause correlations in the sky brightness. Differences between the
spectra of these sources and the average sky brightness will yield
corresponding fluctuations in color (for HST observations, the
``average sky'' is primarily Zodiacal light, for ground-based
observations there is also a strong atmospheric component). 

To constrain possible sources of additional EBL, we compute the
angular autocorrelation function of the unresolved light in the Hubble
Deep Field.  In other words, we constrain additional sources of EBL by
studying the correlation properties of the flux that cannot be
assigned with large statistical significance to individual objects.
Fluctuation analysis of the sky brightness is complementary to direct
measurement of the mean EBL; although the uncertainties in direct
measurement would permit a large EBL component in addition to detected
galaxies, the small amplitude of the sky fluctuations rules out many
plausible candidates for such a component.  For example, we can
statistically test the proposition that previously undetected low
surface brightness galaxies comprise most of the EBL (Vaisanen 1996).
This approach is insensitive to a truly uniform optical background,
but the most plausible source of the EBL is emission from stars, which
are likely to be clustered. At large redshift, we observe these in the
rest-frame UV, where most of the stellar flux is from O and B stars.
Regions of active star formation are extremely clumpy, so we expect
the flux to be strongly correlated.

The approach of studying the correlation properties of the unresolved
optical background was pioneered by Shectman (1974), who used Schmidt
plates to detect optical EBL fluctuations on several arcminute
scales. Shectman detected an EBL power spectrum that was consistent
with clustering of galaxies fainter than $R=18$, as predicted by Gunn
(1965) and Shectman (1973).  Tyson (1988) found evidence for residual
surface brightness fluctuations in deep B band CCD images after
cleaning detected $B<27$mag galaxies and smoothing on a scale of
$6''$.  Cole, Treyer, \& Silk (1992) examined how fluctuation analysis
of deep optical images might constrain faint galaxy populations.
Martin \& Bowyer (1989) detected arcminute-scale fluctuations with a
rocket-borne UV detector, which indicated the presence of a UV
background from galaxies.  Boughn, Saulson, \& Uson (1986) measured
the beam-to-beam variance of the sky flux at $2.2\mu$m to set upper
limits on galaxy counts in the infrared, assuming a model for the
galaxy clustering.  Jenkins \& Reid (1991) also measured sky
fluctuations at $2.2\mu$m to test galaxy evolution models.  Recently,
Kashlinsky et al. (1997) applied a fluctuation analysis to the DIRBE
sky maps to set constraints on the infrared background.  Fluctuations
at zero lag have long been used to constrain the number count
distribution of radio sources (Scheuer 1957; Hewish 1961; Condon 1974;
see, e.g., Fomalont et al. 1988 for a recent result).  Hasinger et
al. (1993) computed limits on the X-ray flux distribution in deep
ROSAT images and several analyses have been performed to measure the
correlation function of the X-ray background, as observed by HEAO 1
(Persic et al. 1989) and Einstein (Barcons \& Fabian 1989).

In this paper we use the Hubble Deep Field to probe the correlation
structure of the night sky after masking all detected objects and
thereby establish constraints on possible sources of the optical
EBL. Section 2 presents estimates of the auto and cross correlation of
the unresolved sky in the Hubble Deep Field.  Section 3 discusses
possible contributions to these measurement from foregrounds and
instrumental systematics.  In Section 4 we use these results to set
stringent limits on the types of additional galaxy populations that
might lie hidden below current detection limits and examine if some of
the measured correlation signal may be caused by weakly clustered
intergalactic light. Section 5 reviews our conclusions.

We quote all HDF magnitudes in the AB system (Oke \& Gunn 1983),
defined as ${\mathrm{AB}}=-2.5\log f_{\nu} -48.60$, where $f_{\nu}$ is
in units of erg/s/cm$^2$/Hz.  Distances are in $h^{-1}$Mpc, where the
Hubble constant is $100h$km s$^{-1}$ Mpc$^{-1}$.

\section{EBL FLUCTUATIONS IN THE HUBBLE DEEP FIELD}

\subsection{Treatment of the HDF Data}

We analyze the drizzled version 2 images of the HDF observed through the
Wide Field cameras (WF) in the F450W, F606W, and F814W filters 
(hereafter we refer to
these bandpasses as $\b450$, $\v606$, and $\i814$, respectively).
Except for corrections to the flat field calibration (see below), we
use the data as reduced and processed by the HDF team (Williams et
al. 1996).  We analyze only the central one-quarter of the area
(roughly $41''\times 41''$) of each drizzled image, to minimize the
possible impact of residual large-scale gradients, and structure in
the dark counts due to fluorescence of the MgF$_2$ window (Burrows et
al. 1995).  [For further details of the HDF, see
http://www.stsci.edu/ftp/science/hdf/hdf.html.]

We apply corrections to the version 2 HDF images to remove large-scale
gradients that are apparent after object-masking and smoothing of the
remaining sky.  Comparison of these features with ratio images formed
from the v2 flat field calibrations and more recent calibration files
(the latter are the calibration files currently used in the WPFC2
pipeline) indicate that the gradients are caused by errors in the v2
calibrations.  To remove these features we smooth the flat field ratio
images (to remove small-scale noise) and multiply them into the HDF
images.  These gradients are quite small, typically $0.5\%$ across an
entire WF image, and are not expected to affect photometry of
individual objects, because algorithms such as FOCAS estimate the sky
level near each object. We remove these gradients because we are
interested in detecting r.m.s. fluctuations in the sky at the level of
$0.1\%$. In section we discuss the possible contribution of residual
flat field uncertainty to the apparent sky fluctuations.

To prepare an object-masked image, we use the HDF version 2 FOCAS
catalog to identify sources with total magnitudes brighter than
$\i814=30$ (Williams et al.  estimate the HDF to be 80\% complete to
$\i814=29$), which includes virtually every object in the catalog.
For comparison, we also derive catalogs from the HDF using the
SExtractor package (Bertin \& Arnouts 1996); we find the mask
structure to be relatively insensitive to the choice of catalog and
choose the available v2 FOCAS catalog to allow others to reproduce our
results.  The same mask, derived from detections in the summed
$\v606+\i814$ image, is used for all filters.  For each object we flag
the pixels that lie within an ellipse that is twice as large as the
area used by FOCAS to compute the ``total magnitude.'' The FOCAS
``total magnitude area'' is always at least twice the area within the
detection isophote, so the masked area is at least four times as large
as the isophotal detection region.  Smoothing of the masked images on
a range of scales does not reveal residual ``doughnuts'' around the
masked objects.  This masking procedure removes $30\%$ of the pixels
in each WF image.  Therefore it does not appear that the sky is
confusion-limited to the isophotal detection threshold of $\sim
27$~mag/arcsec$^2$ that was used for the FOCAS catalog.

\subsection{Correlation Analysis}

The field of view of the WF cameras is well-matched to probing EBL
fluctuations caused by the correlation structure of galaxy
profiles. The WF cameras each have a $80''$ field of view of and
$0.1''$ pixel scale. At redshift $z=1$, $1''$ corresponds to 8.5kpc
($\Omega=1$ and $H_0=50$km/s).  On these angular scales, the
clustering amplitude of $\i814<29$ galaxies is $\omega(1'')\simless
0.1$ (Colley et al. 1996; Villumsen et al. 1997).  This weak
clustering implies that small-scale fluctuations in the EBL will be
dominated by the surface brightness profiles of faint objects rather
than by the clustering among them.

For observed surface brightness $\mu(\xbf)$,
we define the autocorrelation function
\begin{equation}
C(\theta)=\langle \mu(\xbf)\mu(\xbf+\theta)\rangle-\mubar^2,
\end{equation}
where $\mubar=\langle \mu(\xbf) \rangle$.  We often refer to the mean
surface brightness $\mubar$ as the mean ``sky'' brightness (after
masking detected objects, $\mubar$ corresponds to the usual definition
of a ``sky level'').  We use Cartesian coordinates $\xbf$, with
$\theta=|\xbf-\xbf'|$, because the
relative angles in the HDF satisfy $\theta \ll 1$.  The surface
brightness autocorrelation may be written as the product of a
dimensionless autocorrelation function and the square of the mean sky
brightness
\begin{equation}
C(\theta)=\mubar^2 \omega(\theta).
\end{equation}

To prepare an image for correlation analysis, we compute the mean
value of the unmasked pixels and subtract this mean from the image,
set pixels within the object ellipses to zero, and taper the edges of
the image with a cosine bell function. To estimate the autocorrelation
function, we compute the Fourier transform of the image, square the
moduli of the Fourier coefficients (for cross correlations, we
compute the modulus of the products of the images' Fourier
coefficients) and correct
these amplitudes for the effective area of the image after masking and
weighting. We then inverse Fourier transform this two-dimensional
power spectrum to produce the
two-dimensional autocorrelation function.  Averaging over all angles
yields the function $C(\theta)$.  Computation of $C(\theta)$ by direct
summation of the products of the pixel values yields identical results
(but uses significantly more CPU cycles).  Analysis of test images
with known power spectra confirm that we recover the true signal,
after correcting for the integral constraint bias.

The integral constraint bias arises because we estimate the mean
surface brightness $\mubar$ from the image itself. The implicit
assumption that the
ensemble average of $\mu$ is identical to the mean within the image is
equivalent to assuming that there are no fluctuations of the mean
surface brightness between different images.  This bias causes us to
underestimate the autocorrelation by the integral average of the true
autocorrelation function over the area $\Omega$ of the image,
\begin{equation}
\Delta={1\over \Omega^2}\int d^2x\int d^2x'\, \omega_{true}(|\xbf-\xbf'|),
\end{equation}
where $\omega_{true}$ is the true dimensionless autocorrelation
function.

Figure 2 shows the dimensionless autocorrelation function of the
object-masked HDF ``sky'' in each of the three bandpasses.  In each
case, we average the results for the WF2, 3, and 4 fields.
Representative error bars are attached to the $\v606$ curve only and
are estimated from the variation among the three fields and within
bins of width $\delta(\log_{10} \theta)=0.4$.  Near zero lag, the
signal is dominated by photon shot noise over the scale of the $0.1''$
WF pixels.  On scales from $0.15''$ to $8''$, these autocorrelation
functions are well fitted by a power law
\begin{equation}
C(\theta)=\mubar^2 \omega(1'')(\theta/1'')^\gamma, 
\end{equation}
with $\gamma\sim -0.6$. Table 1 summarizes the power law fits and
their uncertainties.  At $1''$, the typical fluctuation is
$\omega(1'')\sim 10^{-6}$, i.e., an r.m.s. correlated fluctuation of
$\sim 0.1\%$ of the mean sky brightness, or $\sim 30.5$~mag/arcsec$^2$ in
$\v606$.

The turnover of $\wtheta$ on scales $\simgreat 10''$ is consistent with the
integral constraint bias that we expect for a power law correlation
function over a field of this size.  Using the power law fits in Table
1, we integrate equation (3) over a $41''\times 41''$ field and find
that the correlation functions in Figure 2 are underestimated by
approximately $\Delta \sim 4\times 10^{-7}$ (note heavy solid bar in
Figure 2).

What could account for the power law shape of these correlations?  In
section 3 we discuss several instrumental systematics and foregrounds
that might contribute to the measured correlations. We note that the
unmasked wings of the point spread function might cause correlations with
similar shape to those plotted in Figure 2.  To explain the shape of
the correlation function with extragalactic sources, one can imagine
an unclustered population of sources of different scale size such that
the sum of their autocorrelations yields a power law.  Alternatively,
the similarity of the power law index, $\gamma\sim -0.6$, to that
observed for galaxy clustering, $\gamma \sim -0.8$, suggests that the
clustering of sources causes the power law shape.  However, because we
have reason to believe that systematics affect this measurement, it is
premature to fit such models.

If we only mask galaxies brighter than $I=26$mag, rather than
$I<30$mag, we find autocorrelations at $\theta < 0.5''$ that are an
order of magnitude above the curves in Figure 2.  This test indicates
that fluctuation analysis of shallower imaging data would be dominated
by the profiles of galaxies detected by the HDF.  In ground-based
imaging, this small-scale correlation structure would be smeared by seeing
and the correlation signal from the profiles of galaxies in the
magnitude range $26<\i814<29$ would be overwhelmingly larger than the
signal measured in the HDF.  In other words, the HDF is the first data
set in which we could have detected a signal as weak as that shown in
Figure 2. One might worry that, just as correlations in shallower
imaging would be dominated by galaxies below the detection threshold,
galaxies just fainter than the HDF limit dominate the observed
correlations; in section 4 we show that this is not the case.

Cross correlations of images in different filters and comparison of
these results with the respective autocorrelations provide a critical
test of the nature of the observed fluctuations.  Dot-dashed curves in
Figure 2 show cross correlations of $\b450$ vs. $\v606$ and $\v606$
vs.  $\i814$.  The cross correlations are shown as
$\omega(\theta)=C_{1\times 2}(\theta) /(\mubar_1 \mubar_2)$, where
$\mubar_1$ and $\mubar_2$ are the mean sky levels.  The photon count
shot noise spike near zero lag is absent in the cross correlations
because this source of noise is uncorrelated between the different
images. Absence of this spike shows that the power law behavior of the
correlations extends down to the smallest scale that we observe.  This
smallest scale is $0.04''$ rather than the $0.1''$ scale of the WF
pixels, thanks to the sub-pixel resolution recovered by drizzling (see
Fruchter \& Hook 1997).

The close agreement between auto and cross correlations indicates that
similar structure is present in different filters, in exposures that
were obtained at different times.  The relative amplitudes are
consistent with a common origin for most of the fluctuations in all
three filters, with some extra signal in the $\b450$ and $\i814$
images that is not present in $\v606$ ($\v606$ also has the highest
signal-to-noise ratio).  The dimensionless correlations in Figure 2
represent fluctuations relative to the mean sky level, thus the good
match between correlations in different filters implies that the color
of the fluctuating component must be close to that of the mean sky. It
is important to remember that, because the mean sky is much brighter
than any possible EBL component, an instrumental systematic that
varies slowly with wavelength could cause correlated structure with
the same color as the mean sky.

We have also computed the autocorrelation function of the color of the
unresolved sky, i.e., the ratio of surface brightness in different
bandpasses (note that this is the ratio of measured surface brightness
{\it before} subtracting the mean sky).  Color correlations of the
$\b450/\v606$ and $\v606/\i814$ ratio images, expressed as
$\omega(\theta)=C_{1/2}(\theta)/(\mubar_1/\mubar_2)^2$, have slightly
smaller amplitude and steeper slope than the auto and cross
correlations.  Color correlations should be relatively less affected
by flatfield uncertainties because any wavelength-{\it independent}
flatfield structure is divided out in forming the ratio image.  We
will report in detail on these results in Paper II (Vogeley 1998).  An
important point is that we expect the amplitude of color correlations
to be smaller than the individual filter autocorrelations by a factor
that strongly depends on the relative color of the sources of the
fluctuations and the mean sky.  Here we note only that the detection
of color correlation of the same order as the auto and cross
correlations argues against a dominant systematic contribution from
flatfield errors.

\section{INSTRUMENTAL SYSTEMATICS AND FOREGROUND SOURCES}

Several instrumental systematics and foreground sources may contribute
to the clustering signal plotted in Figure 2.  These effects include
errors in calibration of the instrument sensitivity, smearing by the
point spread function (PSF) of the telescope, wide-angle scattered
light, and fluctuations in the Zodiacal and Galactic foregrounds.
Because these systematics and foregrounds only add to the measured
clustering signal,
we may use the observed correlations as a firm upper
limit on the true EBL fluctuations; this is the approach that we
follow in Section 4 below.  If we identify an instrumental effect that
contributes to the fluctuations and subtract this contribution, then
we obtain an even stronger constraint on additional sources of the
EBL.  For example, we have already been able to remove some signal by
correction for obvious flatfield errors, and preliminary analysis
clearly indicates a contribution from scattering by the telescope PSF.
In Paper II we provide a detailed analysis of these possible
contributions, with the goal of improving these upper limits.

\subsection{HST/WFPC2 Systematics}

Errors in calibration of spatial variations in the instrument
sensitivity will cause apparent fluctuations in the sky.  As noted
above, smoothing of the object-masked sky revealed large-scale
features in the v2 HDF images, which we removed using better flat
field calibrations. Color correlation analysis removes signal caused
by wavelength-independent flatfield errors, but this does not rule out
a wavelength-dependent component.  Another test is to compare the
two-dimensional autocorrelation structure of the flatfields themselves
with the detected signal.  If the uncertainties in the flatfields are
proportional to the flatfields themselves, then features in their
two-dimensional autocorrelations will show up in the sky correlations.
In addition to cross-correlating the flat field images with the data
directly, we can compare the multipole moments of their
two-dimensional autocorrelation functions.  Preliminary analyses do
not reveal cross correlations or anisotropies that are characteristic
of flat field errors.

Measurement of the multipole moments can also reveal problems caused
by charge transfer efficiency or other defects aligned with the CCD
rows or columns, which cause a quadrapole signal.  Scattered light
from the HST secondary mirror support spider creates an ``X'' pattern
that would yield a large hexadecapole moment.

The point spread function of the WFPC2 has extended wings, which
probably arise from scattering inside the cameras (Krist \& Burrows
1994; Krist 1995; Burrows et al.  1995).  
This PSF is anisotropic and varies with position within each camera.
For our purpose, the most important
effect of the PSF wings
is to scatter some flux well beyond the mask region of an object.
A crude test for this effect is to make an image in which we place a
model of the PSF at the position of each detected object, with
amplitude proportional to the measured magnitude, then mask this image
as we do for the data and cross-correlate with the masked HDF image.
Preliminary results from this and similar tests (such as varying the
mask size) indicate that this effect produces correlations with shape
that is similar to the HDF correlations and could contribute as much as
$10-40\%$ of the observed correlation amplitude.  This may be the
dominant systematic effect on our measurement.  However, it is
difficult to cleanly separate this PSF effect.  Similar effects would
arise if the profiles of detected galaxies extend far beyond the
masked regions (i.e., if the sky is confusion-limited in these
galaxies at some very faint isophote), or if lower surface brightness
objects are clustered with the detected galaxies.

The dark count rate in the WFPC2 varies both spatially and temporally
(Burrows et al. 1995).  These variations are probably caused by
fluorescence of the MgF$_2$ window on the camera.  At fixed epoch, the
dark count rate is constant within the central region of each field,
but declines near the edges of the CCD. The amplitude of the dark
count rate, and therefore the steepness of the roll-off, varies in
time with the cosmic ray flux. To minimize sensitivity to this effect,
we restrict our analysis to the central one-quarter of the area of
each CCD.

The drizzling procedure that was used to combine the HDF exposures
includes corrections for the geometric field distortion in each
camera.  Restricting our analysis to the central area of each field
also minimizes sensitivity to residuals from this distortion.

\subsection{Foreground Sources}

The motion of the Earth relative to the source of the Zodiacal light
causes fluctuations in this foreground to be decorrelated between
WFPC2 exposures. The Zodiacal light is Solar flux that is
backscattered from a layer of dust that extends to roughly 3
A.U. (Dermott et al. 1996; Reach et al. 1996).  In the time it takes
for HST to orbit the Earth (96 minutes), the Earth's motion around the
Sun changes the HST's line of sight through this dust layer.  During
the month of December (when the HDF was observed), the line of sight
through a screen at a distance of 3 A.U. towards a fixed extrasolar
target at $12^h37^m$ $+62^{\circ}13'$ (the position of the HDF)
changes by $\sim 2'$.  The field of view of a WF camera is $\sim
1.25'$, therefore sky brightness fluctuations due to the Zodiacal
foreground become decorrelated if we examine fluctuations by cross
correlating WFPC2 exposures that are separated in time by at least
the duration of one orbit.  Exposures in different filters were
separated by many orbits, thus cross correlations of images in
different filters should be unaffected by ZL fluctuations.  The good
match of auto and cross correlations indicates that ZL fluctuations do
not strongly affect any of our measurements.

The next brightest foreground is Galactic cirrus, which is presumably
the reflection of starlight from high-latitude dust.  The HDF field
was chosen to lie at a minimum in the IRAS $100\mu$m maps, thus we
expect this Galactic foreground to be much smaller than average.
Guhatakurta \& Tyson (1989) measured the color of the Galactic cirrus
and found that it is $\sim 1^m$ redder in $B_J-R$ than either the faint blue
galaxies or the structure with surface brightness fainter than $\sim
30$ mag/arcsec$^2$ seen on scales $>6''$ by Tyson (1988).  This color
is similarly too red to match the color implied by the auto and cross
correlations of the HDF sky (within a few tenths of a magnitude of the
Zodiacal light color).  However, because little is know about
correlations of this foreground on scales of a few arcseconds, we
cannot exclude this possibility.

\section{CONSTRAINTS ON SOURCES OF THE EBL}

\subsection{Galaxy Populations}

Is there a population of sources that (1) would not have been directly
detected in the HDF, (2) would contribute significant surface
brightness to the EBL, and (3) would not overproduce fluctuations in
the EBL?  If we specify the distribution of fluxes, surface brightness
profiles, and angular clustering of a proposed undetected population,
then we can compute the effect of these sources on fluctuations in the
EBL.  Using the measured correlations of the object-masked sky in the
HDF as an upper limit on the true EBL fluctuations, we then test
whether the predicted fluctuations are allowed by the measured sky
correlations.
 
The autocorrelation of the surface brightness from a population of
sources is the sum of contributions from the correlation structure
within the profiles and from clustering among the sources.  For a
population of sources with identical apparent surface brightness
profiles $\mu_{gal}(\theta)$, the autocorrelation function is
\begin{equation}
C_{galaxies}(\theta)=\overline{n} c_{gal}(\theta) + \overline{n}^2
\int d^2x \, c_{gal}(\theta-\xbf)\omega_{clust}(\xbf).
\end{equation}
The first term is from the correlations of flux within the object
profiles; $c_{gal}$ is the convolution of a galaxy profile with
itself,
\begin{equation}
c_{gal}(\thbf)=\int d^2x\, \mu_{gal}(\xbf) \mu_{gal}(\xbf+\thbf).
\end{equation}
The second term is from clustering among the galaxies, with angular
two-point correlation $\omega_{clust}(\theta)$.

In this subsection we assume that $\omega_{clust}=0$, so that we can
constrain possible EBL sources on the basis of their profiles
alone. The addition of source clustering would increase the predicted
EBL fluctuations and strengthen the constraints on these models. In
the next subsection we examine the other extreme, a clustered diffuse
EBL source.

Because it does not account for absorption of flux from one galaxy by
dust in another, equation (5) is exact only in the limit where the
sources are transparent or sparsely distributed.  If galaxies have
finite optical depth then we require more sources to produce fixed
optical EBL.  Another effect of absorption is that clustering of
sources causes a smaller increase in the EBL fluctuations because
nearby galaxies tend to shield one another.  A complete accounting for
these effects requires specification of the redshift distribution,
spectral energy distribution, and extinction law for the sources.  We
will examine such detailed models in Paper III.

Here we consider only simple phenomenological source models, in which
unclustered sources have exponential surface brightness profiles with
identical apparent central surface brightness $\mu_0$ and apparent
angular scale length $h$.  Such a population contributes total surface
brightness
\begin{equation}
\mubar_{pop}=\overline{n}f=E \mubar_{detect},
\end{equation}
where $\overline{n}$ is the projected number density, $f=2\pi\mu_0
h^2$ is the total flux of each source, and $E$ is the ratio of this
extra surface brightness to the surface brightness in detected
galaxies to the HDF limit, $\mubar_{detect}$ (here we define
$\mubar_{detect}$, which is the same as $\mubar_{galaxies}$ defined in
section 1, to avoid confusion with the additional source population).

To compute the EBL fluctuations from this source population, we first
compute the convolution of an exponential disk with itself,
\begin{equation}
	c_{exp}(\theta)=c_{exp}(0) p(\theta/h) = 
	\left({\pi \over 2} \mu_0^2 h^2 \right ) 
	\left({2\over \pi} \int d^2u\, e^{-|{\bf u}|}
	e^{-\left\vert{\bf u}+ \theta /h  \right \vert}\right),
\end{equation}
which defines a dimensionless profile function $p(\theta/h)$ with
$p(0)=1$.  The autocorrelation of the population is
\begin{equation}
C_{pop}(\theta)=\overline{n}{\pi \over 2}\mu_0^2 h^2 p(\theta/h)
= {\overline{n}f^2 \over 8\pi h^2} p(\theta /h).
\end{equation}

The power law fits to the HDF correlations (Table 1) yield a
constraint on the EBL fluctuations caused by this population,
\begin{equation}
 C_{pop}(\theta)<C_{obs}(1'')\theta^{\gamma}
\end{equation}
for all $\theta$.  Combining the equations above, we derive a
constraint on the the $\v606$ band central surface brightness $\mu_0$
and angular scale length $h$ of these sources,
\begin{equation}
\mu_0 < 6.3\times 10^{-22} (h/1'')^{-0.6} E^{-1} \,
{\mathrm{erg/s/cm^2/sr/Hz}}.
\end{equation}

If we define an effective area $\pi h^2$ for each source, then the
covering factor (the average number of objects along a random line of
sight) is of order
\begin{equation}
\chi=\overline{n}\pi h^2={E\mubar_{detect} \over 2 \mu_0}.
\end{equation} 

Figure 3 shows the resulting constraints on the $\v606$ band central
surface brightness $\mu_0$ and scale size $h$ of a population of
identical unclustered exponential disks that contributes mean surface
brightness equal to that of detected galaxies ($E=1$). The region
below the solid line satisfies the constraint of equation (11).  Above
this line the sources' autocorrelation function would exceed this
upper limit for some range of $\theta$.  The right-hand axis indicates
the covering factor $\chi$ for different $\mu_0$.  Above the dashed
line at the top of this figure lie objects directly detected in the
HDF using FOCAS with an isophotal threshold of $\sim
27$~mag/arcsec$^2$.  For a smaller contribution to the EBL, i.e., for
$E<1$, we relax the $\mu_0$ limit by $2.5\log E$ and change the
covering factor by a factor $E^{-1}$.  In the limit $E\simless 0.1$,
the allowed region abuts the detected region.

Two examples of populations that contribute
$\mubar_{pop}=\mubar_{detect}$ and marginally satisfy the correlation
constraint illustrate how very different such a population would be
from the HDF detections.  For $h=0.1''$ a population of extremely
faint, total magnitude $\v606=32.4$, sources with $\mu_0=29.4$
~mag/arcsec$^2$ is marginally allowed.  The number of such objects,
$n=1.3\times 10^9$deg$^{-2}$, is $\sim 600$ times the total number per
deg$^2$ of detected galaxies.  For $h=1.0''$, the correlations allow a
population of $\v606=28.9$mag sources, with
$\mu_0=30.9$~mag/arcsec$^2$.  For comparison, the typical scale size
of $\v606=29$mag detected galaxies is $\simless 0.2''$.  The projected
number density of the additional $\v606=28.9$mag population would be
20 times the total number of detected objects. On the righthand axis
of this plot we note the covering factor that corresponds to each
choice of $\mu_0$ and $h$. For the former example, this covering
factor is $\chi \sim 3$, for the latter $\chi \sim 13$.  Such large
covering factors imply that these objects might be detected as QSO
absorption line systems and/or from reddening of background objects.

For comparison to these models, note that the most extended low
surface brightness galaxy seen to date, Malin 1, has an extrapolated
disk central surface brightness of $\mu_0=26$~mag/arcsec$^2$ in V and
scale size $h=82h^{-1}_{75}$kpc (Bothun et al. 1987). At $z=0.5$ this
would be dimmed to roughly $28$~mag/arcsec$^2$ in I and would have an
apparent scale size $h=17''$.  We obtain relatively poor constraints
on sources with $h\simgreat 10''$ because the integral constraint bias
is comparable to the correlation function on these scales (see Table
1); the $41''\times 41''$ field that we analyze is too small to
accurately measure larger scale fluctuations.  However, for $E=1$,
objects like Malin 1 clearly would be ruled out in Fig. 3 for a
reasonable extrapolation of the measured correlation function.

Thus, to make a large contribution to the EBL, the correlation
constraint requires that this additional population and the detected
galaxies form an extremely bimodal distribution in surface brightness.
The undetected sources must have very low central surface brightness
and be confusion-limited on the sky.  Some galaxies certainly lie
within the region between the FOCAS detections and the ``allowed''
region for extra sources of the EBL.  As shown in Section 1,
extrapolation of the number counts would place some objects here but
they would not contribute very much surface brightness to the EBL.

Extrapolation of the detected galaxy counts adds little to the EBL,
but would these fainter galaxies cause detectable correlations?
No. Following equation (9), a population of galaxies fainter than
$m_{lim}$ with number per magnitude distribution 
$N(m)=d\overline{n}/dm$ and identical
angular scale size $h$ has zero-lag autocorrelation signal
\begin{equation}
C_{pop}(0)={1\over 8\pi h^2}\int_{mlim}^{\infty} dm\, N(m)
\left[10^{-0.8(m+48.60)} \right] {\mathrm{(erg/s/cm^2/sr/Hz)^2}}.
\end{equation}
If we extrapolate the HDF V-band counts from $V=29.5$mag to zero flux
with logarithmic count slope $\alpha=0.2$ and scale size $h=0.2''$,
then $C(0)_{pop}=2.2\times 10^{-43}$(erg/s/cm$^2$/sr/Hz)$^2$, which
translates to a dimensionless correlation
$\omega_{pop}(0)=C_{pop}(0)/\mubar^2=2.2\times 10^{-7}$.  Comparison
to Figure 2 shows that this signal is two orders of magnitude below
the measured correlations.  Thus, although galaxies brighter than the
HDF detection limit would dominate the clustering signal (see section
2), galaxies just fainter than this limit have little effect on the
observed correlations.

\subsection{Weakly Clustered Diffuse Light}

So far, we have ignored clustering among the undetected objects. If
the undetected sources are clustered, then the solid line in Figure 3
moves down towards even lower allowed central surface brightness.  In
the limit of point-like objects, the autocorrelation function of a
population (eq.[5]) reduces to
\begin{equation}
	C_{pop}(\theta)=\overline{n}f^2 \delta(\theta)
	+\overline{n}^2f^2\omega_{clust}(\theta).
\end{equation}
This must be convolved with the PSF to determine the observed
autocorrelation.  If we set $\overline{n}f=E\mubar_{detect}$ then we
obtain a simple constraint on the $1''$ angular clustering amplitude
of the sources,
\begin{equation}
	\omega_{clust}(1'') < {C_{obs}(1'') \over (E\mubar_{detect})^2}.
\end{equation}
This constraint applies for all scales
if we make the additional assumption that these objects have power law
angular correlations, $\omega_{pop}(\theta)\propto \theta^{\gamma}$
with $\gamma\sim -0.6$, similar to that fit to the HDF sky correlations.

On scales larger than the PSF size, the inequality in equation (15) is
equivalent to a constraint on a clustered diffuse EBL component with
$\mubar_{diffuse}=E\mubar_{detect}$.  Figure 4 shows the constraints
on combinations of $\mubar_{diffuse}$ and $\omega_{diffuse}(1'')$ from
the HDF correlations.  For $E=1$, the upper bound on angular
clustering of diffuse light is $\omega_{diffuse}(1'') < 8\times
10^{-3}$. This $1''$ clustering amplitude is an order of magnitude
smaller than the clustering measured for galaxies as faint as $I<29$
in the HDF (Colley et al. 1996; Villumsen et al. 1997).  Thus, to
contribute as much surface brightness as the detected galaxies, these
undetected objects must be not only extremely faint and numerous, but
also uniformly distributed to exquisite precision.

Apart from {\it imagined} sources of additional EBL, several
observations of galaxy clusters detect diffuse light or intergalactic
stars that would contribute to the amplitude and clustering of the EBL
if similar surface brightness is associated with most galaxies.  From
R band imaging of Abell 2029, Uson, Boughn, \& Kuhn (1991) infer that
10\% of the light in this cluster is in a diffuse component.  Using
WFPC2 imaging of the Virgo cluster, Ferguson, Tanvir, \& Von Hippel
(1996) find a population of intergalactic stars that could contribute
total flux equal to 10\% of that in galaxies.  Theuns \& Warren (1996)
detect candidate planetary nebulae in Fornax and infer an
intergalactic stellar population that contributes as much as 40\% of
the cluster light. M\'endez et al. (1997) also find candidates for
intergalactic planetary nebulae in Virgo.  Tyson, Kochanski, \&
Dell'Antonio (1997) detect diffuse light in CL0024+1654 that comprises
$15\%$ of the light in the central 100kpc of this cluster.

If diffuse intergalactic light makes a fractional contribution to the
EBL (relative to $\mubar_{detect}$ from galaxy counts) that is similar
to the intergalactic surface brightness detected in clusters, then it
could cause EBL fluctuations close to those measured in the HDF sky.
The constraints on diffuse intergalactic light also follow from
equation (15) if the diffuse light is clustered in similar fashion to
faint galaxies, with a power law correlation function
$C_{diff}(\theta)\sim \mubar_{diff}^2 \theta^{-0.6}$.  Let us suppose
that the clustering amplitude of the diffuse light is comparable to
that of faint galaxies.  The small arrow in Figure 4 indicates the
level of diffuse light that contributes an additional
$\mubar_{diffuse}=0.1 \mubar_{detect}$ (i.e., similar to the $10\%$
contribution of intergalactic stars in Virgo).  The EBL correlations
from this source would match the HDF correlations if this light has
the same clustering amplitude as $R<23$mag galaxies, $\omega(1'')\sim 1$
(Couch, Jurcevic, \& Boyle 1993).  Because $75\%$ of the light from
detected galaxies comes from $\i814<23$ galaxies, it seems plausible
that much of the diffuse light would be associated with galaxies in
this same magnitude interval, and be gravitationally clustered with
similar amplitude.  Note that the contribution of any such a source to
the EBL fluctuations would leave even less room in the upper limits on
sky correlations for a discrete source population.

\subsection{Is the HDF Typical?}

A remaining question is whether the HDF is a typical field for the
purpose of probing fluctuations in the EBL.  The answer depends on the
redshift and angular size of the hypothesized sources of undetected
EBL. The $80''\times 80''$ field of view of each WF camera limits the
utility of the HDF for studying fluctuations on scales much larger
than $10''$. This angular coverage also limits the range of apparent
magnitude of the galaxies that the field includes.  Because this
observation was designed to be an unbiased probe of the high-redshift
universe, the field was deliberately chosen to avoid bright foreground
galaxies ($V< 22$mag) which would fill much of the WFPC2 field of
view.

This avoidance of bright galaxies raises the concern that the HDF is
biased against detection of EBL fluctuations from previously
undetected sources at redshift $z\simless 0.3$ that are clustered with
bright galaxies.  [A similar bias arises in the direct EBL detection
method of Bernstein et al., because they use WFPC2 imaging to measure
the total sky brightness.]  However, it seems extremely contrived to
envisage a population of sources that is not detected at $z>0.3$ through
fluctuation analysis, is not directly detected at $z<0.3$ in deep
ground-based imaging, and yet contributes a large fraction of the
optical background.

Another issue is variation of EBL fluctuations from field to field.
The total imaged area of the three WF cameras is 5.3 arcmin$^2$, of
which we analyze only the central 1.3 arcmin$^2$.  If additional
sources of EBL are distributed similarly to the detected galaxies,
then expected fluctuations in galaxy counts provides a rough estimate
of uncertainty in the mean EBL over this area.  The Poisson
fluctuation in counts of galaxies with magnitude $26<V<29.5$ would be
5\% over a 1.3 arcmin$^2$ area. For the magnitude range $23<V<26$, the
expected fluctuation rises to 12\%.  Galaxy clustering does not
significantly increase this lower bound on the field-to-field
fluctuations; although the field of view spans only a few hundred kpc
in angle, these magnitude intervals include galaxies over many
hundreds of comoving Mpc in distance. We conclude that the
field-to-field variation in the amplitude of the HDF correlations is
likely to be caused by systematic effects rather than true
fluctuations in the EBL.

\section{CONCLUSIONS}

Fluctuations in the object-masked sky in the HDF provide a strong,
albeit indirect, constraint on the mean optical EBL. If the HDF
provides a fair probe of the extragalactic sky between detected
galaxies, then we conclude that this observation and other deep
imaging surveys have already detected the sources of the majority of
the optical EBL. The mean optical EBL is, at most, a few tens of
percent above the mean surface brightness from detected galaxies.  At
$\lambda=8100\AA$, a generous allowance for up to 50\% additional EBL
suggests $7.8\times 10^{-6}\le \nu I_{\nu} < 1.2\times
10^{-5}$erg/s/cm$^2$/sr.  This mean level lies at the lower end of the
uncertainty range of the measurement by Bernstein et al.  ($\nu
I_{\nu}=2.1\pm 1.2\times 10^{-5}$erg/s/cm$^2$/sr, after including
$V<23$mag galaxies) and well below previous upper limits.

We infer this mean EBL level because no plausible sources of
additional extragalactic surface brightness can satisfy the multiple
constraints that these sources (1) have not already been detected, (2)
contribute total surface brightness comparable to the surface
brightness contributed by detected galaxies, and (3) do not produce
EBL fluctuations larger than the upper limits set by correlations in
the object-masked HDF.  Figure 3 shows that these constraints admit
only extremely low surface brightness objects that must be so extended
and numerous as to be confusion-limited on the sky. The parameters of
simple phenomenological models that meet these constraints would make
such a population disjoint from the parameter space of all detected
objects.  Extrapolation of the HDF galaxy number counts to infinitely
faint limits would add at most a few percent to the mean EBL.  Figure
4 shows that a diffuse component that is as large as the surface
brightness from detected galaxies would require a $V$-band angular
clustering amplitude of $\omega(1'')<8\times 10^{-3}$, an order of
magnitude less clustered than the faintest observed galaxies.
Fluctuations in the EBL from intergalactic stars as seen in galaxy
clusters could be as large as the upper limits on the fluctuations in
the HDF, but would contribute only incrementally to the mean EBL.

In addition to constraining the plausible level of the optical EBL,
these upper limits on small-scale sky fluctuations in the HDF provide
a test for any proposed model for faint galaxy populations.  We plan
to test galaxy population models that are more detailed than the
simple phenomenological model described in Section 4 (Paper III).  We
invite others to suggest possible models, compute the predicted
autocorrelation function of sources below the detection limits, and
compare with these limits from the HDF.  These correlations are also
important for attempts to detect weak lensing signals in deep optical
imaging (Van Waerbeke et al. 1997; Refregier et al. 1997).

As we emphasize throughout this paper, the measured correlations of
the object-masked sky in the HDF should be treated as upper limits on
the true EBL fluctuations.  The wings of the WFPC2 point spread
function almost certainly contribute to the measured fluctuations.
Although there are reasons to exclude flat fielding errors or Galactic
light as the principal sources of the fluctuations, we cannot rule out
some contribution from these effects.  To improve these upper limits,
we are examining these systematics and foregrounds in further detail
(Paper II).

The HDF provides the best constraint to date on small-scale EBL
fluctuations because its superior resolution allows detection of faint
galaxy populations that would be confusion-limited in ground-based
imaging. In section 2, we discuss how fluctuation analysis of
shallower HST or ground-based images would be dominated on small
scales by the profiles of galaxies that only the HDF detects.
However, ground-based imaging is required to study fluctuations on
angular scales much larger than several arcseconds, where access to
larger collecting area and accurate control of systematics outweigh
the HST's advantages of high resolution and lower sky background. If
we measure object-masked sky correlations in deep ground-based images,
we can now use the HDF galaxy counts and profiles to model and
subtract the surface brightness correlations that are caused by
galaxies that are fainter than the ground-based detection limits, but
that were detected by the HDF. Fluctuation analysis of deep
ground-based imaging will allow statistical tests for diffuse
intergalactic light on larger angular scales and for low surface
brightness companions to galaxies that were too bright to be included
in the HDF (see section 4c).

Future HST observations will allow measurement of EBL fluctuations in
independent fields and at fainter surface brightness levels than the
HDF. The planned HDF South (October 1998) will reach to similar depth
and allow us to test if the HDF North is typical.  The Advanced Camera
for Surveys (scheduled for installation on HST in 1999) will have
superior sensitivity and better-controlled systematic uncertainties
than WFPC2, as well as a $200'' \times 200''$ contiguous field of
view, thus allowing measurement of surface brightness fluctuations at
fainter levels and somewhat larger scales.  The proposed GTO program
for ACS focuses on deep imaging of galaxy clusters.  Fluctuation
analysis of the ACS cluster fields will be an important test for
sources of diffuse intergalactic surface brightness.  In the
$2-10\mu$m infrared, the proposed Next Generation Space Telescope
might detect statistical fluctuations in the EBL from the first
generation of stars (e.g., Haiman \& Loeb 1997) even if such objects
evade direct detection.

\acknowledgements

We thank R. Williams and the entire HDF team, whose vision and effort
made this project possible. 
We acknowledge discussions with and many helpful comments from 
A. Babul,
R. Bernstein,
S. Casertano, 
A. Connolly,
H. Ferguson, 
A. Fruchter,
M. Geller, 
E. Groth,
R. Lupton,
P. Madau,
J. Peebles,
A. Refregier,
P. Stockman, 
M. Strauss,
A. Szalay,
T. Tyson,
and D. Wilkinson.
We also thank R. Bernstein and L. Pozzetti for
providing results in advance of publication.
This paper was initiated during a visit to the Aspen Center for
Physics.
Support for this work was provided by NASA through grants
HF-01078.01-94A and AR-05812.01-94A from the Space Telescope Science
Institute, which is operated by AURA, Inc.  under NASA contract
NAS5-26555,

\newpage

\begin{table}
\label{tab1}
\caption{Autocorrelation and Cross Correlation Power Law Fits
\tablenotemark{a} 
}
\vskip 24pt
\begin{tabular}{ccccc}
\tableline
Filter & $\omega(1'')$ & $\gamma$ & $\mubar$
\tablenotemark{b} & $\Delta$
\tablenotemark{c} \\
\tableline
$\b450$ & $3.58\, (\pm 0.69)\times 10^{-6}$ & 
$-0.80\, (\pm 0.16)$ & $6.62\times 10^{-19}$ &
$4.14\times 10^{-7}$ \\
$\v606$ & $1.99\, (\pm 0.38)\times 10^{-6}$ & 
$-0.61\, (\pm 0.14)$ & $1.00\times 10^{-18}$ &
$3.67\times 10^{-7}$ \\
$\i814$ & $2.99\, (\pm 0.59)\times 10^{-6}$ & 
$-0.59\, (\pm 0.13)$ & $1.39\times 10^{-18}$ &
$5.84\times 10^{-7}$ \\
$\b450 \times \v606$ & $1.96\, (\pm 0.39)\times 10^{-6}$ & 
$-0.62\, (\pm 0.15)$ & ... &
$3.53\times 10^{-7}$ \\
$\v606 \times \i814$ & $2.14\, (\pm 0.46)\times 10^{-6}$ & 
$-0.53\, (\pm 0.14)$ & ... &
$4.90\times 10^{-7}$ \\
\end{tabular} 
\tablenotetext{a}{The observed correlation function 
over $0.15''<\theta<8''$ is fit by
$C(\theta)=\mubar^2 \omega(1'') (\theta/1'')^{\gamma}$. 
}
\tablenotetext{b}{Mean of object-masked sky in erg/s/cm$^2$/sr/Hz.
Note that $10^{-18}$ erg/s/cm$^2$/sr/Hz $=22.94$ AB mag/arcsec$^2$.
}
\tablenotetext{c}{Integral constraint bias for 
a $41''\times 41''$ field, estimated using these fits (See eq.[3]).
}
\end{table}


\newpage

\begin{figure}
\figurenum{1}
\plotone{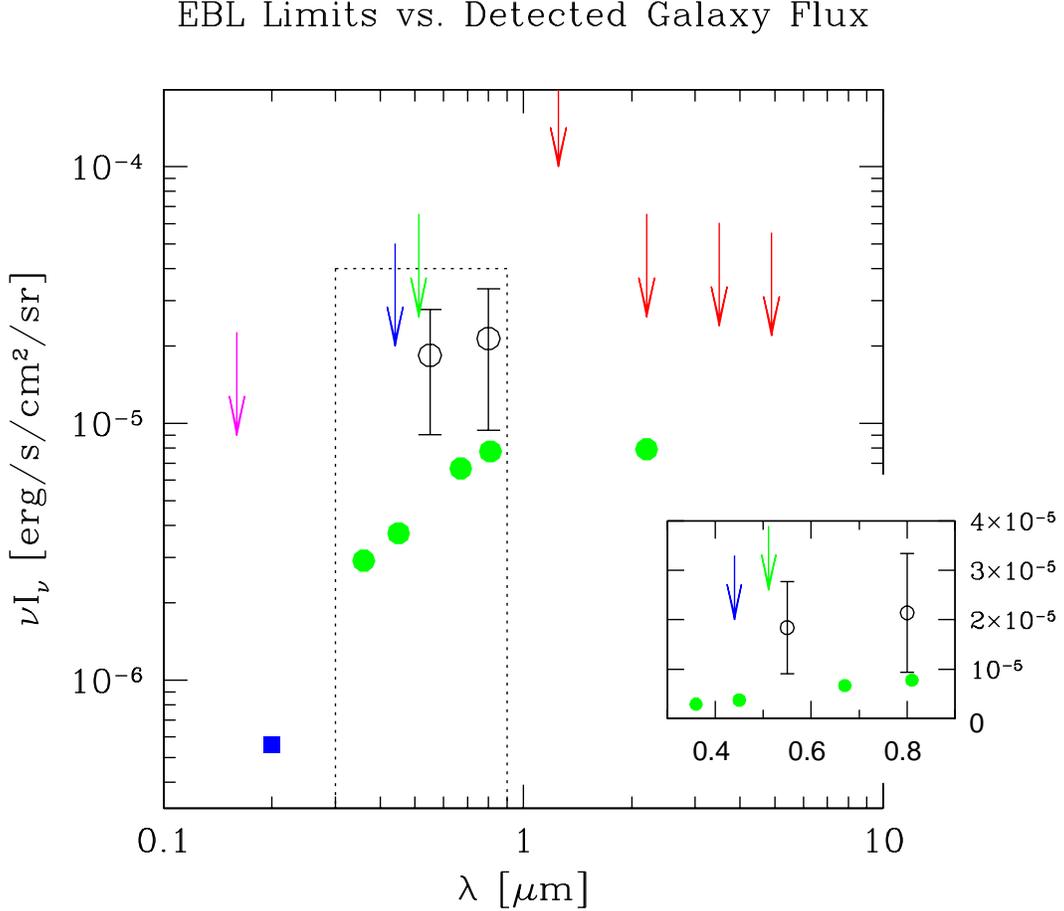}
\caption{Observational constraints on the mean EBL. 
Inset box replots the dotted region (optical limits only) using linear axes.
Solid symbols are
lower limits from the average surface brightness contributed by
galaxies detected in deep imaging surveys: the HDF, other optical
surveys, and K-band (solid circles; Pozzetti et al. 1997), 
UV (square; Milliard et al. 1992).  Arrows indicate
upper limits on the extragalactic contribution to the night sky
brightness in the UV from Paresce (1990), in the optical from
Toller (1983) and Dube et al. (1977, 1979), and in the IR from
Hauser (1996) (left to right, respectively).  Open circles indicate a
detection of the mean EBL (Bernstein 1997) with $1\sigma$
uncertainties.  The detection level includes a small adjustment for
flux from bright galaxies ($V<23$mag) that were excluded from this
measurement.  The surface brightness in AB magnitudes that corresponds
to $\nu I_{\nu}= 10^{-5}$ erg/s/cm$^2$/sr at $\lambda=5500\AA$ is 27.3
mag/arcsec$^2$.}
\end{figure}

\begin{figure}
\figurenum{2}
\plotone{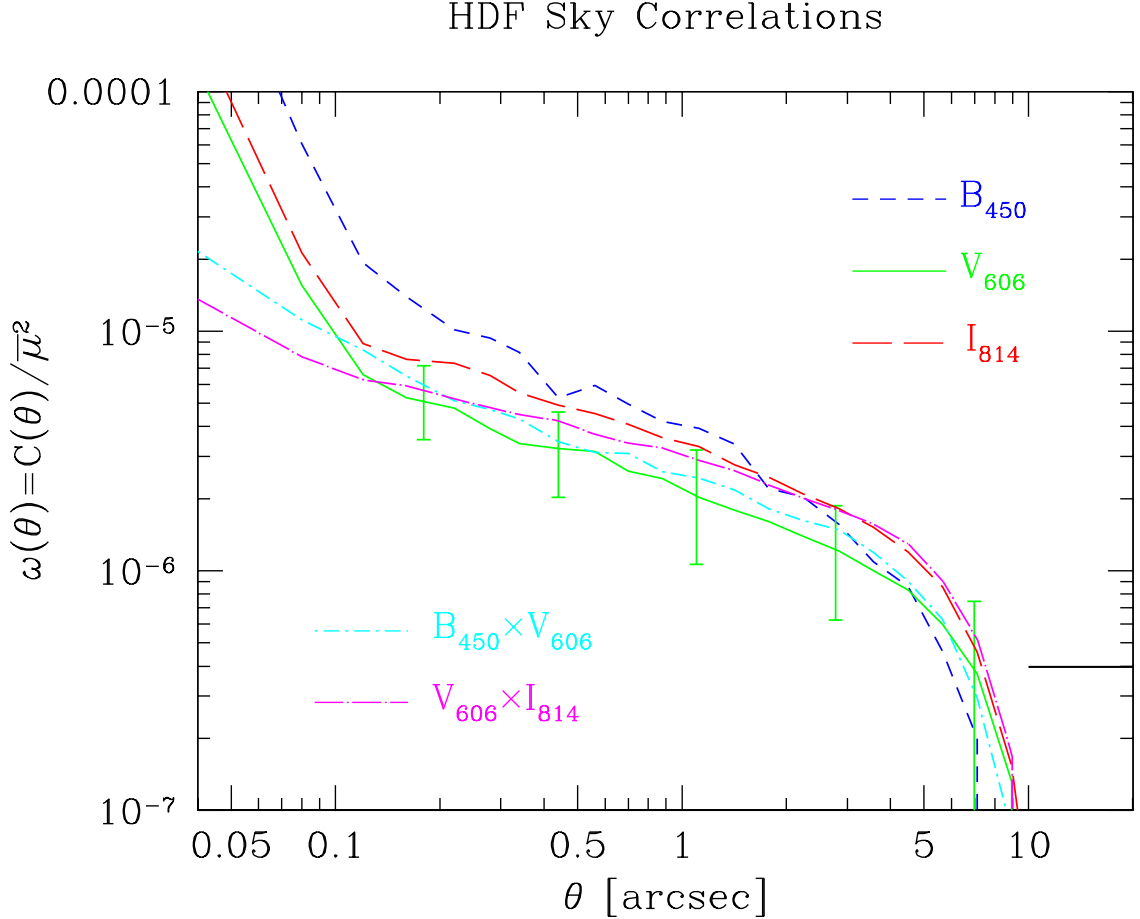}
\caption{Autocorrelations of ``sky'' pixels in the central $41''
\times 41''$ of the HDF images in the $\b450$, $\v606$, and $\i814$
bandpasses, and cross correlations of $\b450\times \v606$ and
$\v606\times \i814$.  These curves are averages over the WF2, 3, and 4
fields.  Error bars on the $\v606$ curve are from the variance among
the fields (which dominates this uncertainty) and within bins of width
$\delta(\log_{10}\theta)=0.4$. Other curves have similar uncertainty.  Shot
noise dominates the autocorrelations for $\theta<0.1''$.  Both auto
and cross correlations are well-fitted by power laws from $0.15''$ to
$8''$ (see Table 1 for fitting parameters).  The drop-off at large
$\theta$ is an artifact of the integral constraint bias on the
correlation function for a finite area of sky. The heavy solid bar at
$4\times 10^{-7}$ marks the typical amount by which we underestimate
the correlations due to this effect.  Cross correlation removes the
shot noise and confirms that roughly the same fluctuations are seen in
different filters.  Several sources of instrumental systematics and
astronomical foregrounds may contribute to this measured correlation
signal. Thus, we treat this measurement as an upper limit on
small-scale fluctuations in the EBL.}
\end{figure}

\begin{figure}
\figurenum{3}
\plotone{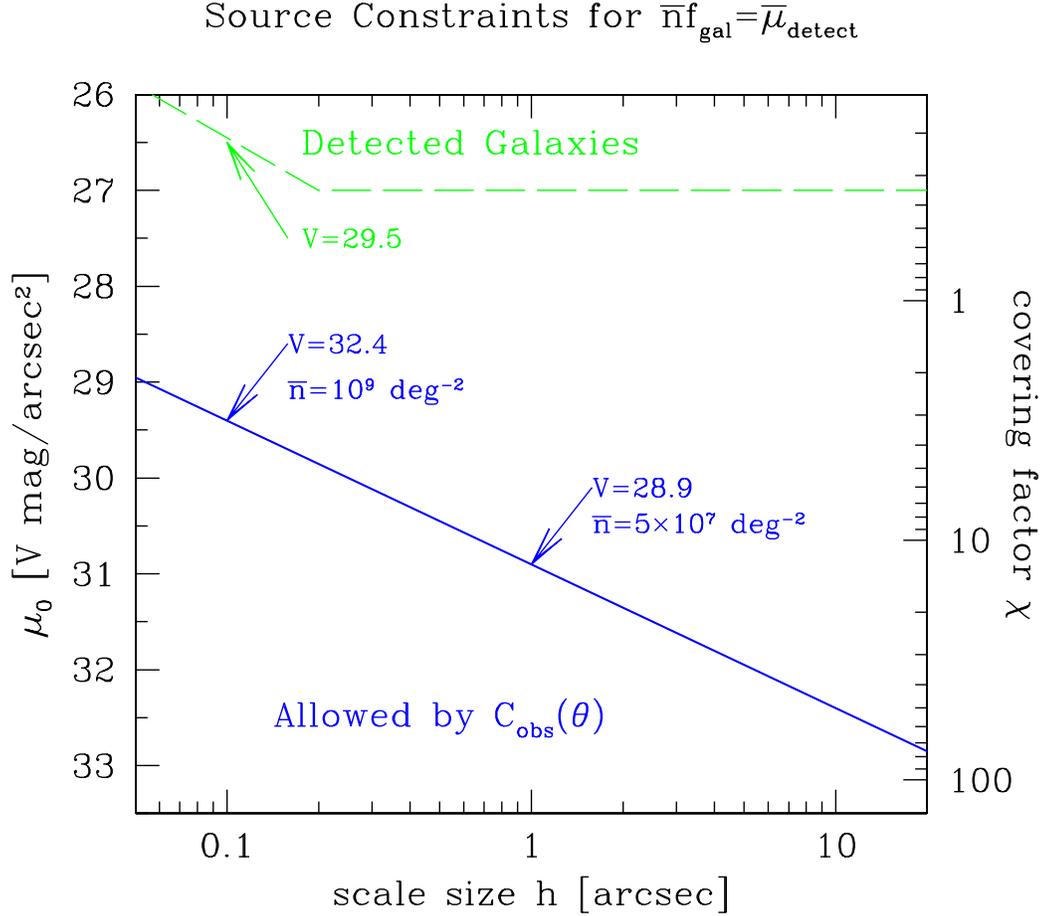}
\caption{Correlation constraints on a population of identical
exponential disks, assuming no clustering among the objects.  Using
the power-law correlation signal in Figure 2 as an upper limit on the
true fluctuations, here we explore the allowed range of $h$ and
$\mu_0$ in the $\v606$ band.  The surface density of objects is set so
that the EBL contributed by these sources is equal to that in detected
galaxies.  A population with central surface brightness $\mu_0$ and
scale size $h$ that lies above the solid line would produce
correlations larger than observed.  Below this line, the correlation
constraint is satisfied, but the covering factor on the sky
(right-hand axis) is quite large; any population with
$\mu_0>29.5$mag/arcsec$^2$ would completely cover the sky $>3$ times.
Indicated are the total magnitudes and surface density of two
marginally-allowed populations.  The region above the dashed line is
the locus of galaxies detected to the limits of the HDF.  Some
galaxies may exist between the dashed box and solid line without
causing excessive EBL correlations; they simply cannot contribute very
much total surface brightness to the EBL.}
\end{figure}

\begin{figure}
\figurenum{4}
\plotone{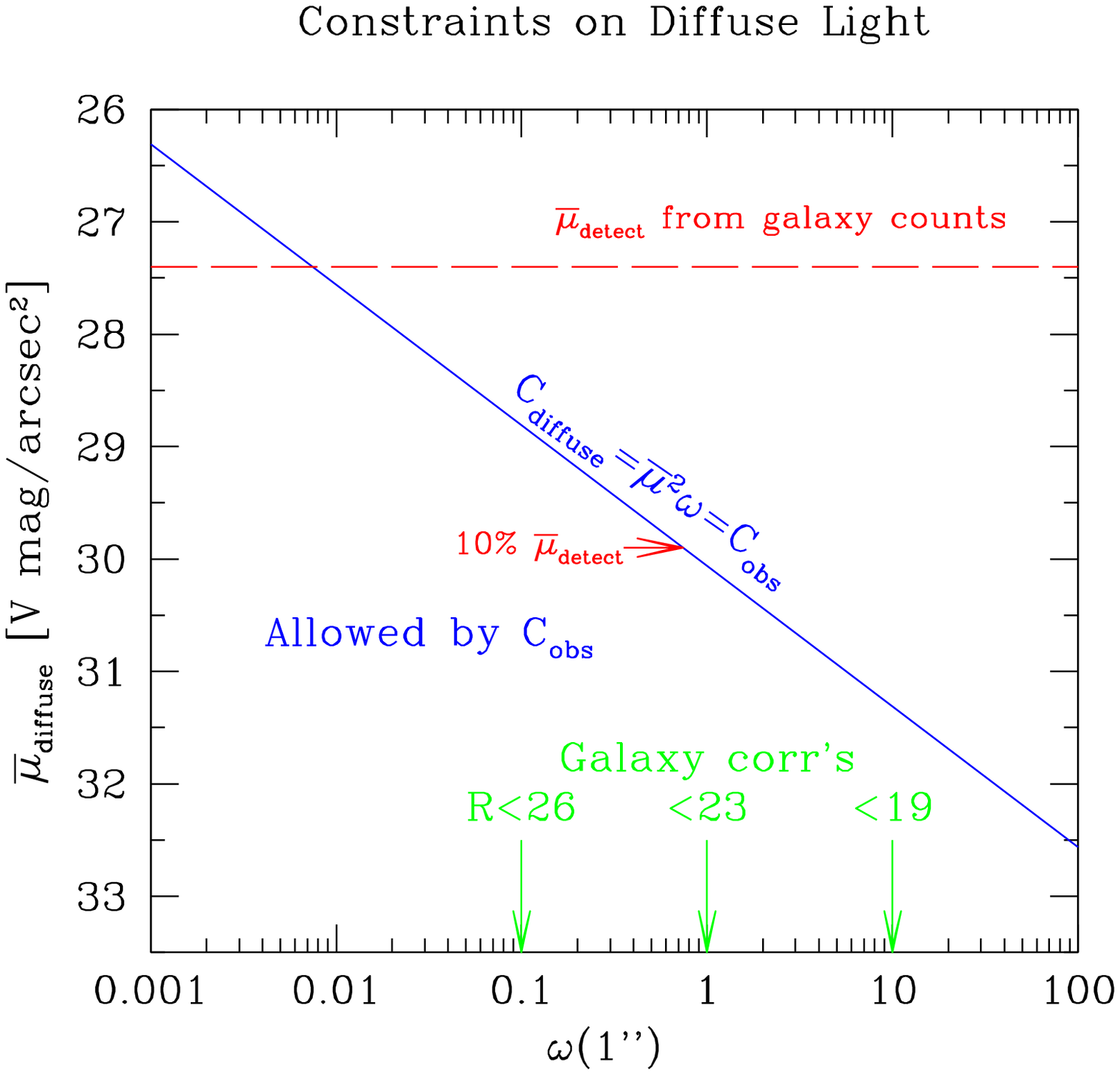}
\caption{Constraints on clustered diffuse light with mean surface
brightness $\mubar$ and $1''$ clustering amplitude $\omega(1'')$,
assuming that the angular correlation function of the diffuse light
has the same slope as the measured $\omega(\theta)$.  The allowed
region in the $\mubar-\omega(1'')$ plane lies below the solid line.
To contribute surface brightness equal to that in detected galaxies
(dashed line), the diffuse light would have to be nearly uniform.  A
small fractional contribution to the EBL, similar to the fraction of
diffuse light in clusters, might be allowed if this light has angular
clustering similar to faint galaxies (note arrows at bottom of this
figure).  The observed correlations marginally allow diffuse light
with surface brightness that is $\sim 10\%$ that of detected galaxies
if the correlation amplitude is similar to $R<23$mag galaxies.  Note
that 75\% of the resolved EBL comes from galaxies with $I \simless
23$mag.  Such an EBL component would account for the observed
fluctuations but would add only incrementally to the total mean EBL.}
\end{figure}


\begin{references}


\reference{} Barcons, X., \& Fabian, A. C. 1989, \mnras, 237, 119

\reference{} Bernstein, R. A. 1997, Ph.D. thesis, California Institute
of Technology

\reference{} Bernstein, R. A., Freedman, W. L., \& Madore, B. F. 1998,
in preparation

\reference{} Bertin, E., \& Arnouts, S. 1996, A\&A Suppl., 117, 393

\reference{} Bothun, G. D., Impey, C. D., Malin, D. F., \& Mould, J. R.
1987, \aj, 94, 23

\reference{} Boughn, S. P., \& Kuhn, J. R. 1986, \apj, 309, 33

\reference{} Boughn, S. P., Saulson, P. R., \& Uson, J. M. 1986, \apj,
301, 17

\reference{} Burrows, C. J., et al. 1995, Wide Field and Planetary Camera 2
Instrument Handbook, STScI publication

\reference{} Cole, S., Treyer, M., \& Silk, J. 1992, \apj, 385, 9

\reference{} Colley, W. N., Rhoads, J. E., Ostriker, J. P., \&
Spergel, D. N. 1996, \apj, 43, 63

\reference{} Condon, J. J. 1974, \apj, 188, 279

\reference{} Connolly, A. J., Szalay, A. S., Dickinson, M., Subbarao,
M. U., \& Brunner, R. J. 1997, \apjl, 486, L11

\reference{} Couch, W. J., Jurcevic, J. S., \& Boyle, B. J. 1993,
\mnras, 260, 241

\reference{} Dalcanton, J. J. 1997, \apj, in press

\reference{} Dermott, S. F., Jayaraman, S., Xu, Y. L., Grogan, K., \&
Gustafson, A. S. 1996, in Unveiling the Cosmic Infrared Background,
ed. E. Dwek, AIP Conf. Proc. 348, 25

\reference{} Disney, M. J. 1976, Nature, 263, 573

\reference{} Djorgovski, S., Soifer, B. T., Pahre, M. A., Larkin,
J. E., Smith, J. D., Neugebauer, G., Smail, I. Matthews, K., Hogg,
D. W., Blandford, R. D., Cohen, J., Harrison, W., \& Nelson, J. 1995,
\apjl, 438, L13

\reference{} Dube, R. R., Wickes, W. C., \& Wilkinson, D. T. 1977,
\apj, 215, L51

\reference{} Dube, R. R., Wickes, W. C., \& Wilkinson, D. T. 1979,
\apj, 232, 333

\reference{} Fall, S. M., Charlot, S. \& Pei, Y. C. 1996, \apjl, 464, L1

\reference{} Ferguson, H., Tanvir, N.  \& Von Hippel, T. 1996, \baas,
189, 80.06

\reference{} Fomalont, E. B., Kellerman, K. I., Anderson, M. C., Weistrop,
D., Wall, J. V., Windhorst, R. A., \& Kristian, J. A. 1988, \aj, 96, 1187

\reference{} Fruchter, A. S., \& Hook, R. N. 1997, in Applications of
Digital Image Processing XX, ed. A. Tescher, Proc. S.P.I.E. vol 3164,
in press (preprint astro-ph/9708242)

\reference{} Guhathakurta, P., \& Tyson, J. A. 1989, \apj, 346, 773

\reference{} Gunn, J. E. 1965, Ph.D. thesis, California Institute of
Technology

\reference{} Haiman, Z., \& Loeb, A. 1997, \apj, 483, 21

\reference{} Hasinger, G., Burg, R., Giacconi, R., Hartner, G., Schmidt, M.,
Truemper, J., \& Zamorani, G. 1993, \aap, 275, 1; erratum: 291, 348

\reference{} Hauser, M. G. 1996, in Unveiling the Cosmic Infrared Background,
ed. E. Dwek, AIP Conf. Proc. 348, 11

\reference{} Hewish, A. 1961, \mnras, 123, 167

\reference{} Kashlinsky, A., Mather, J. C., \& Odenwald, S. 1997,
\apjl, 473, L9

\reference{} Krist, J. E. 1995, in Calibrating the Hubble Space Telescope:
Post Servicing Mission, eds. A. Koratkar \& C. Leitherer, 311

\reference{} Krist, J. E., \& Burrows, C. J. 1994, WFPC2 Instrument Science
Report 94-01, STScI

\reference{} Lillie, C. F. 1968, Ph.D. thesis, University of
Wisconsin

\reference{} Madau, P., Ferguson, H. C., Dickinson, M. E., Giavalisco,
M., Steidel, C. C., \& Fruchter, A. 1996, \mnras, 283, 1388

\reference{} Madau, P., Pozzetti, L., \& Dickinson, M. 1997, \apj, submitted
(preprint astro-ph/9708218)


\reference{} Martin, C., \& Bowyer, S. 1989, \apj, 338, 677

\reference{} Mattila, K. 1990, in The Galactic and Extragalactic
Background Radiation, Proc. IAU Symp. 139, eds. S. Bowyer \&
C. Leinert (Dordrecht: Kluwer), 257

\reference{} McGaugh, S. S., Bothun, G. D., \& Schombert, J. M. 1995,
\aj, 110, 573

\reference{} M\'endez, R. H., Guerrero, M. A., Freeman, K. C.,
Arnaboldi, M., Kudritzki, R. P., Hopp, U., Capaccioli, M., \& Ford, H. 1997,
\apjl, in press (preprint astro-ph/9710179)

\reference{} Milliard, B., Donas, J., Laget, M., Armand, C., \&
Vuillemin, A. 1992, \aap, 257, 24

\reference{} Oke, J. B., \& Gunn, J. E. 1983, \apj, 266, 713

\reference{} Paresce, F. 1990, in The Galactic and Extragalactic
Background Radiation, Proc. IAU Symp. 139, eds. S. Bowyer \&
C. Leinert (Dordrecht: Kluwer), 307

\reference{} Partridge, B., \& Peebles, P. J. E. 1967a, \apj, 147, 868

\reference{} Partridge, B., \& Peebles, P. J. E. 1967b, \apj, 148, 377

\reference{} Persic, M., de Zotti, G., Boldt, E. A., Marshall, F. E., 
Danese, L., Francheschini, A., \& Palumbo, G. G. C. 1989, \apjl, 336, L47


\reference{} Pozzetti, L., Madau, P., Zamorani, G., Ferguson, H. C.,
\& Bruzual, G. A. 1997, \mnras, submitted

\reference{} Reach, W. T., Franz, B. A., Kelsall, T., \& Weiland, J. L. 
1996, in Unveiling the Cosmic Infrared Background,
ed. E. Dwek, AIP Conf. Proc. 348, 37

\reference{} Refregier, A., et al. 1997, in preparation

\reference{} Roach, F. E., \& Smith, L. L. 1968, Geophys. J., 15, 227

\reference{} Scheuer, P. A. G. 1957, Proc. Camb. Phil. Soc., 53, 764

\reference{} Shectman, S. A. 1973, \apj, 179, 681

\reference{} Shectman, S. A. 1974, \apj, 188, 233

\reference{} Spinrad, H., \& Stone, R. P. S. 1978, \apj, 226, 609

\reference{} Toller, G. N. 1983, \apjl, 266, L79

\reference{} Theuns, T. \& Warren, S. J. 1996,
\mnras, 284, 11

\reference{} Tyson, J. A. 1988, \aj, 96, 1

\reference{} Tyson, J. A. 1995, in Extragalactic Background Radiation,
eds. D. Calzetti, M. Livio, \& P. Madau (Cambridge University Press), 103

\reference{} Tyson, J. A., Kochanski, \& Dell'Antonio, I. 1997,
\apj, submitted

\reference{} Uson, J. M., Boughn, S. P., \& Kuhn, J. R. 1991, \apj,
369, 46

\reference{} Van Waerbeke, L., Mellier, Y., Schneider, P., Fort, B., \&
Mathez, G. 1997, \aap, 317, 303

\reference{} Vaisanen, P. 1996, \aap, 315, 21

\reference{} Villumsen, J. V., Freudling, W., \& da Costa, L. N. 1997,
\apj, 481, 578

\reference{} Vogeley, M. S. 1998, in preparation (Papers II, III)

\reference{} Whitrow, G. J., \& Yallop, B. D. 1965, \mnras, 130, 31

\reference{} Williams, R. E., Blacker, B., Dickinson, M.,
Dixon, W. V. D., Ferguson, H. C., Fruchter, A. S., Giavalisco,
M., Gilliland, R. L., Heyer, I., Katsanis, R., Levay, Z., Lucas, R. A.,
McElroy, D. B., Petro, L., Postman, M., Adorf, H. M., \& Hook,
R. N. 1996, \aj, 112, 1335

\end{references}
\end{document}